\documentstyle [12pt] {article}
\title{A CLOSE CORRELATION BETWEEN THIRD KEPLER LAW AND TITIUS-BODE RULE}

\author{Vladan Pankovi\'c$^{\ast,\sharp}$,
Aleksandar-Meda Radakovi\'c$^\sharp$\\
$^\ast$Department of Physics, Faculty of Sciences, 21000 Novi
Sad,\\ Trg Dositeja Obradovi\'ca 4. , Serbia, vdpan@neobee.net \\
$^\sharp$Gimnazija, 22320 Indjija, Trg Slobode 2a, Serbia \\}

\date {}
\begin{document}
\maketitle

\vspace {0.1cm} PACS number:  96.35.-j \vspace {0.1cm}

\begin {abstract}
In this work we present a close correlation between third Kepler
law and Titius-Bode empirical rule. Concretely, we demonstrate
that third Kepler law, or, corresponding equilibrium condition
between centrifugal and Newtonian gravitational force, implies
that planet orbital momentum becomes effectively a function of the
planet distance as unique variable and vice versa. Then,
approximation of the planet distance by its first order Taylor
expansion over planet orbital momentum holds an exponential form
corresponding to Titius-Bode rule. In this way it is not necessary
postulate exponential form of the planet distance (as it has been
done by Scardigli) but only discrete values of its argument.
Physically, it simply means that, in the linear approximation,
"quantized" planets orbital momentums do a geometrical
progression.
\end {abstract}

In this work we shall present a close correlation between third
Kepler law and Titius-Bode empirical rule [1], [2]. Concretely, we
shall demonstrate that third Kepler law, or, corresponding
equilibrium condition between centrifugal and Newtonian
gravitational force, implies that planet orbital momentum becomes
effectively a function of the planet distance as unique variable
and vice versa. Then, approximation of the planet distance by its
first order Taylor expansion over planet orbital momentum holds an
exponential form corresponding to Titius-Bode rule. In this way it
is not necessary postulate exponential form of the planet distance
(as it has been done by Scardigli [1]) but only discrete values of
its argument. Physically, it simply means that, in the linear
approximation, "quantized" planets orbital momentums do a
geometrical progression.

Consider well-known situation when a relatively small physical
system, e.g. a planet in Sun system, stablely rotates, by means of
Newtonian gravitational force, along a circumference about
central, massive system, e.g. Sun. It corresponds to equilibrium
between centrifugal and Newtonian gravitational force, i.e. to
expression
\begin {equation}
      \frac {mv^{2}}{R} = \frac {GmM}{R^{2}}                                            .
\end {equation}
Here $m$ represents the planet mass, $M$ - Sun mass, $R$ - planet
orbit radius or distance between planet and Sun, $v=\frac {2\pi
R}{T}$ - planet speed, $T$ - revolution period and $G$ - Newtonian
gravitational constant.

After division of (1) by $\frac {m}{R}$ and use of the definition
of $v$ as $\frac {2\pi R}{T}$, it simply follows
\begin {equation}
      \frac {T^{2}}{R^{3}}= \frac {4\pi ^{2}}{GM}
\end {equation}
that represents the remarkable third Kepler law.

Expression (1) represents generally speaking a functional
dependence between two variables, $v$ and $R$. Value of one of
these variables, e.g. $v$, can be chosen, in principle, quite
arbitrarily and then, according to (1), there is a practically
continuous spectrum of the values of other variable, e.g. $R$.

But, as it is well-known too, according to famous Titius-Bode [1]
empirical rule in Richardson form [1], [2]
\begin {equation}
       R_{n} = R_{1} \exp[2\alpha (n-1)] \hspace{1cm}   {\rm for} \hspace{1cm} n=1, 2, 3,... .
\end {equation}
it seems that planets orbits are discretized, i.e. "quantized",
where $R_{1}$ represents Mercury orbit radius, $R_{2}$ - Venus
orbit radius, etc. and $2\alpha = 0.53707$ corresponding parameter
characteristic for Sun system. If Titius-Bode rule does not
represent a coincidence only, its existence implies an additional,
"quantization" physical condition (dynamical or kinematical) {\it
strongly} different from equilibrium condition (1) or Kepler law
(2), as it is presented in [1], [2] etc.

For example Scardigli [2], in an incomplete analogy with Bohr
momentum quatization, postulated
\begin {equation}
      J = mvR =  mS \exp[\alpha n] \hspace{1cm}   {\rm for} \hspace{1cm} n=1, 2, 3,... .
\end {equation}
where $J=mvR$ represents a planet orbital momentum and $S$ - an
additional parameter. Then (1) and (4) represent the equations
system with two variables $R$ and $v$, whose unique solution
predict "quantized" form of $R$ equivalent to (3). Explanation of
this postulate Scardigli gives by introduction of a more complex
theory, i.e. a more accurate, Schr$\ddot{o}$dinger-like dynamics
of the global structure of planetary system.

It can be observed that condition (1) can be simply, i.e. by
multiplication with $mR^{3}$, transformed in the following
expression
\begin {equation}
      J^{2}= m^{2}v^{2}R^{2} = (Gm^{2}M) R
\end {equation}
that implies
\begin {equation}
      R = (Gm^{2}M) ^{-1}J^{2}                                    .
\end {equation}

Expressions (5) and (6) are very interesting. Namely, according to
its definition, $J=mvR$, planet orbital momentum $J$ represents
the function of two practically independent variables, planet
circumference radius $R$, and planet speed $v$. However, according
to equilibrium condition (1), $J$ becomes function of only one
variable $R$ (5), and vice versa (6).

Now, we shall approximate (6) by its first order Taylor expansion
within a small vicinity $\Delta J = J - J_{0}$ of an orbital
momentum value $J_{0}$. It yields

      \[R \simeq (Gm^{2}M) ^{-1}J^{2}_{0}+ 2(Gm^{2}M) ^{-1}J_{0}\Delta J
      =\]
\begin {equation}
      (Gm^{2}M) ^{-1}J^{2}_{0} + 2(Gm^{2}M) ^{-1}J^{2}_{0}\frac {\Delta
      J}{J_{0}}= R_{0}+ 2R_{0}\frac {\Delta J}{J_{0}}= R_{0}(1 + 2\frac {\Delta
J}{J_{0}})
\end {equation}
where $R_{0}=(Gm^{2}M) ^{-1}J^{2}_{0}$. Since $(1 + 2\frac {\Delta
J}{J_{0}})$ can be considered as first order Taylor expansion of
the expression $\exp[2\frac {\Delta J}{J_{0}}]$ we shall suggest a
more accurate expression
\begin {equation}
      R= R_{0}\exp[2\frac {\Delta J}{J_{0}}] .
\end {equation}
It and (1) imply
\begin {equation}
      v = v_{0}\exp[-\frac {\Delta J}{J_{0}}]
\end {equation}
where $v_{0} = (GM)^{\frac {1}{2}}R^{-\frac {1}{2}}_{0}$. Also,
according to definition of $J$ and (8), (9), it follows formally
\begin {equation}
     J = J_{0}\exp[\frac {\Delta J}{J_{0}}]
\end {equation}
that is satisfied in the first order approximation, i.e. Taylor
expansion, where J0 = mv0R0.
   Suppose now that there is a discrete planet orbital momentum series corresponding to (10)
\begin {equation}
     J_{n}= J_{n-1}\exp[\frac {\Delta J_{n}}{J_{n-1}}]  \hspace{1cm}   {\rm for} \hspace{1cm} n=2, 3,... . .
\end {equation}
It, obviously, can be transformed in
\begin {equation}
     J_{n} = J_{1}\exp[\frac {\Delta J_{n}}{J_{n-1}}+\frac {\Delta J_{n-1}}{J_{n-2}}+… +\frac {\Delta J_{2}}{J_{1}}]  \hspace{0.5cm}   {\rm for} \hspace{0.5cm} n=1, 2, 3,... ..
\end {equation}
    Suppose additionally
\begin {equation}
     \frac {\Delta J_{n}}{J_{n-1}}= \alpha \simeq const \hspace{1cm}   {\rm for} \hspace{1cm} n=2, 3,... .
\end {equation}
where $\Delta J_{n} = J_{n}-J_{n-1}$ for   $n=2, 3, …$ .It means
that given planet orbital  momentum series $J_{1}, J_{2}, … ,
J_{n}$  represents a geometrical progression with coefficient $1+
\alpha$. Then (12) turns out in
\begin {equation}
     J_{n}= J_{1}\exp[(n-1) \alpha] \hspace{1cm}   {\rm for} \hspace{1cm} n=2, 3,... .    .
\end {equation}
It implies
\begin {equation}
      R_{n}= R_{1}\exp[2(n-1) \alpha] \hspace{1cm}   {\rm for} \hspace{1cm} n=2, 3,... .
\end {equation}
and
\begin {equation}
      v_{n} = v_{1}\exp[-(n-1)\alpha] \hspace{1cm}   {\rm for} \hspace{1cm} n=2, 3,... .  .
\end {equation}
   Obviously, (15) has the form equivalent to Richardson form of Titius-Bode empirical rule (3).
    In this way it is proved that Titius-Bode rule follows directly from equilibrium condition (1) or third Kepler law (2) under an additional, weak, "quantization" physical condition (13) which, physically, simply means that, in the linear approximation, "quantized" planets orbital momentums do a geometrical progression. (Given condition is weak in the sense that it is not necessary postulate exponential form of $R$ but only discrete values of the argument of exponential function.)

\vspace{1.5cm}

{\Large \bf References}

\begin {itemize}

\item [[1]] M. M. Nieto, {\it The Titius-Bode Law of Planetary Distances: its History and Theory} (Pergamon Press, Oxford, 1972)
\item [[2]] F. Scardigli, {\it A Quantum-like Description of the Planetary Systems}, gr-qc/0507046, and references therein

\end {itemize}

\end {document}